\documentclass{sig-alternate-2013}

\begin{document}
%
\conferenceinfo{PIR'14,Privacy-Preserving IR: When Information Retrieval Meets Privacy and Security, SIGIR 2014 Workshop,} {July 11th, 2014, Gold Coast, Australia} 
\CopyrightYear{2014} 
\clubpenalty=10000 
\widowpenalty = 10000

\title{Privacy-Preserving Important Passage Retrieval}
%
%
%
%
%

\numberofauthors{1} 
%
\author{
%
%
\alignauthor{Lu\'{i}s Marujo$^{1,2,3}$, Jos\'{e} Port\^{e}lo$^{2,3}$, David Martins de Matos$^{2,3}$, Jo\~{a}o P. Neto$^{2,3}$, \\
      Anatole Gershman$^{1}$, Jaime Carbonell$^{1}$, Isabel Trancoso$^{2,3}$, Bhiksha Raj$^{1}$ \\}
\affaddr{$^1$ Language Technologies Institute, Carnegie Mellon University, Pittsburgh, PA, USA \\
         $^2$ Instituto Superior T\'{e}cnico, Lisboa, Portugal; \\
         $^3$ INESC-ID, Lisboa, Portugal \\}
\email{\{lmarujo,anatoleg,jgc,bhiksha\}@cs.cmu.edu, \{jose.portelo,david.matos,joao.neto,isabel.trancoso\}@inesc-id.pt}
}

\maketitle
\begin{abstract}
State-of-the-art important passage retrieval methods obtain very good results, but do not take into account privacy issues. In this paper, we present a privacy preserving method that relies on creating secure representations of documents. Our approach allows for third parties to retrieve important passages from documents without learning anything regarding their content. We use a hashing scheme known as Secure Binary Embeddings to convert a key phrase and bag-of-words representation to bit strings in a way that allows the computation of approximate distances, instead of exact ones. Experiments show that our secure system yield similar results to its non-private counterpart on both clean text and noisy speech recognized text.
\end{abstract}




\section{Introduction}
\label{sec:intro}
{\em Important Passage Retrieval} (IPR) is the problem of extracting the most important passages in a body of text. By ``important'', we mean those passages that capture most of the key information the text is attempting to convey. Of the various solutions proposed, state-of-the-art solutions for IPR based on {\em centrality} achieve excellent results \cite{Ribeiro&Marujo:2013}. 

A potential problem to the deployment of such methods is that they usually assume that the input data are of public domain. However, this data may come from social network profiles, medical records or other private documents, and their owners may not want to, or even be allowed to share it with third parties. Consider the scenario where a company has millions of classified documents. The company needs to retrieve the most important passages from those documents, but lacks the computational power or know-how to do so. At the same time, they can not give access to the documents to a third party with such capabilities because they may contain sensitive information. As a result, the company must {\em obfuscate} their own data before sending it to the third party, a requirement that is seemingly at odds with the objective of extracting important passages from it.

In this paper, we propose a new {\em privacy-preserving} technique for IPR based on Secure Binary Embeddings (SBE) \cite{SBE} that enables exactly this -- it provides a mechanism for obfuscating the data, while still achieving near state-of-the-art performance in IPR. 

SBEs are a form of locality-sensitive hashing which convert data arrays such as bag-of-words vectors to obfuscated bit strings through a combination of random projections followed by banded quantization. The method has information theoretic guarantees of security, ensuring that the original data cannot be recovered from the bit strings. At the same time, they also provide a mechanism for {\em locally} computing distances between vectors that are close to one another without revealing the global geometry of the data, consequently enabling tasks such as IPR. This is possible because, unlike other hashing methods which require exact matches for performing classification tasks, SBEs allows for a near-exact matching: the hashes can be used to estimate the distances between vectors that are very close, but provably provide no information whatsoever about the distance between vectors that are farther apart.
The usefulness of SBE has already been shown for implementing a privacy-preserving speaker verification system \cite{PPSV_SBE} yielding promising results.

The remainder of the paper is structured as follows. In Section \ref{sec:related_work} we briefly present some related work regarding Important Passage Retrieval and privacy-preserving methods in IR. In Section \ref{sec:IPR} we detail the two stages of the important passage retrieval technique. Section \ref{sec:SBE} presents the method for obtaining a secure representation method. We describe our approach to privacy-preserving important passage retrieval in Section \ref{sec:secure_passage_retrieval}. Section \ref{sec:experiments} describes the dataset used and illustrates our approach with some experiments. Finally, we present some conclusions and plans for future work.

\section{Related Work}
\label{sec:related_work}

\subsection{Important Passage Retrieval}
Text and speech information sources influence the complexity of the important passage retrieval approaches differently. For textual passage retrieval, it is common to use complex information, such as syntactic \cite{vanderwende:et:al:2007}, semantic \cite{tucker:jones:2005}, and discourse information \cite{uzeda:et:al:2010}, either to assess relevance or reduce the length of the output. However, speech important passage retrieval approaches have an extra layer of complexity, caused by speech-related issues like recognition errors or disfluencies. As a result, it is useful to use speech-specific information (e.g.: acoustic/prosodic features \cite{maskey:hirschberg:2005}, recognition confidence scores \cite{zechner:waibel:2000}), or by improving both the assessment of relevance and the intelligibility of automatic speech recognizer transcriptions (by using related information \cite{ribeiro:matos:2013}). These problems not only increase the difficulty in determining the salient information, but also constrain the applicability of passage retrieval techniques to speech passage retrieval. Nevertheless, shallow text summarization approaches such as Latent Semantic Analysis (LSA) \cite{landauer:dumais:1997} and Maximal Marginal Relevance (MMR) \cite{carbonell:goldstein:1998} seem to achieve performances comparable to the ones using specific speech-related features \cite{penn:zhu:2008}. In addition, discourse features start to gain some importance in speech retrieval \cite{maskey:hirschberg:2005,zhang:chan:fung:2010}.

Closely related to the important passage retrieval used by this work are approaches using the unsupervised key phrase extraction methods. These methods are used to reinforce passage retrieval \cite{zha2002generic,wan2007towards,litvak:last:2008,riedhammer:favre:tur:2010,Sipos:2012}. Namely, they propose the use of key phrases to summarize news articles \cite{litvak:last:2008} and meetings \cite{riedhammer:favre:tur:2010}. In \cite{litvak:last:2008}, the authors explored both supervised and unsupervised methods with a limited set of features to extract key phrases as a first step towards important passage retrieval. Furthermore, the important passage retrieval used in this work adapts the centrality retrieval model, which plays an important role in the whole process. This kind of model adaptation is explored in \cite{riedhammer:favre:tur:2010}, where the first stage of their method consists in a simple key phrase extraction step, based on part-of-speech patterns; then, these key phrases are used to define the relevance and redundancy components of a MMR summarization model.


Most of the IPR methods could be easily adapted to be secure using the method described in Section \ref{sec:SBE}. We opted to use the KP-Centrality method described in the next section because it has the current state-of-the-art IPR method.

\subsection{Privacy Preserving Methods}
In this work, we focused on creating a method to perform important passage retrieval keeping the information in the original documents private. There is a large body of literature on important passage retrieval and privacy preserving or secure methods. To the best of our knowledge, the combination of both research lines has not been explored yet. However, there are some recent works combining information retrieval and privacy. Most of these works use data encryption \cite{Jiang:2007,MurugesanEtAl:2010,Jiang:2011,Lu:2014} to transfer the data in a secure way. This does not solve our problem because the content of the document would be decrypted by the retrieval method and therefore it would not remain confidential to the retrieval method. Another alternative secure information retrieval methodology is to obfuscate queries, which hides user topical intention \cite{Pang:2012}, but does not secure documents content.

In many areas the interest in privacy-preserving methods where two or more parties are involved and they wish to jointly perform a given operation without disclosing their private information is not new, and several techniques such as Garbled Circuits \cite{yao1982protocols}, Homomorphic Encryption \cite{paillier1999public}, Lo-cality-Sensitive Hashing \cite{indyk1998approximate} have been introduced. However, they all have limitations regarding the Important Passage Retrieval task we wish to address. Until recently Garbled Circuits were extremely inefficient to use due to several intrinsic issues, and even now it is difficult to adapt them when the computation of non-linear operations is required. Solutions to many of these problems have been developed, such as performing offline computation of the oblivious transfers, using shorter ciphers, evaluating XOR gates for 'free', etc.~\cite{bellare2013efficient}. Systems based on Homomorphic Encryption techniques introduce substantial amounts of computational overhead and usually require extremely long amounts of time to evaluate any function of interest. The Locality-Sensitive Hashing technique allows for near-exact match detection between data points, but does not provide any actual notion of distance, leading to degradation of performance in some applications. As a result, we decided to consider Secure Binary Embeddings \cite{SBE} as the data privacy method for our approach, as it does not show any of the disadvantages mentioned above for the task at hand. We describe this technique in depth in Section \ref{sec:SBE}.

\section{Important Passage Retrieval}
\label{sec:IPR}
To determine the most important sentences of an information source, we used the KP-Centrality model \cite{Ribeiro&Marujo:2013}. We chose this model for its adaptability to different types of information sources (e.g., text, audio and video) and state-of-the-art performance. It is based on the notion of combining key phrases with support sets. A support set is a group of the most semantically related passages. These semantic passages are selected using heuristics based on the passage order method \cite{ribeiro:matos:2011}. This type of heuristic explore the structure of the input source to partition the candidate passages to be included in the support set in two subsets: the ones closer to the passage associated with the support set under construction and the ones further apart. These heuristics use a permutation, $d^i_1, d^i_2, \cdots, d^i_{N-1}$, of the distances of the passages $s_k$ to the passage $p_i$, related to the support set under construction, with $d^i_k = dist(s_k, p_i)\text{, }1 \le k \le N-1$, where $N$ is the number of passages, corresponding to the order of occurrence of passages $s_k$ in the input source. The metric that is normally used is the cosine distance.

\begin{figure*}[!t]
\centering
\includegraphics[width=1.6\columnwidth]{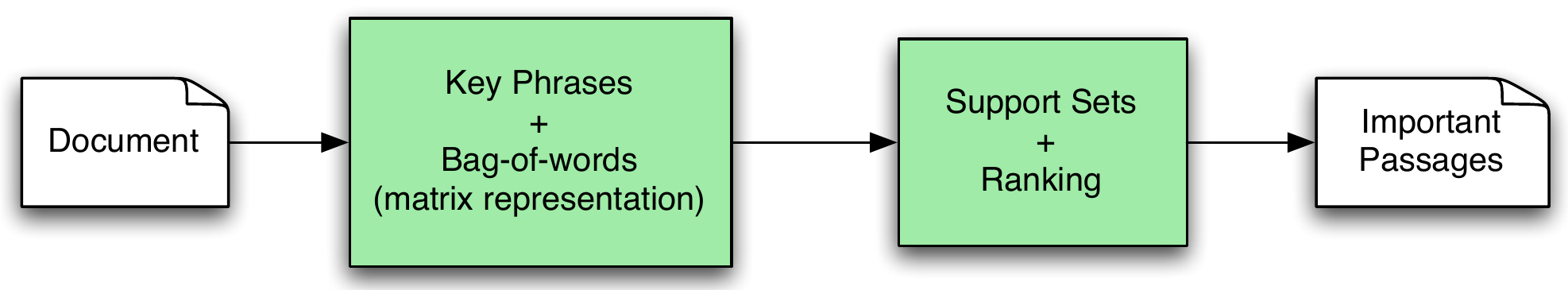}
\caption{Flowchart of the Important Passage Retrieval method.}
\label{figure:ipr_diagram}
\end{figure*}

The KP-Centrality method consists of two steps, as illustrated in Figure \ref{figure:ipr_diagram}.
First, it extracts key phrases using a supervised approach \cite{Marujo_LREC_2012} and combines them with a bag-of-words model in a compact matrix representation, given by:
\begin{equation}
\begin{bmatrix}
  w(t_1,p_1) & \dots & w(t_1,p_N) & w(t_1,k_1) & \dots & w(t_1,k_M)\\
  \vdots & & & & & \vdots\\
  w(t_T,p_1) & \dots & w(t_T,p_N) & w(t_T,k_1) & \dots & w(t_T,k_M)\\
\end{bmatrix},
\label{eq:kpcentrality}
\end{equation}
where $w$ is a function of the number of occurrences of term $t_i$ in passage $p_j$ or key phrase $k_l$, $T$ is the number of terms and $M$ is the number of key phrases. Then, using a segmented information source $I \triangleq p_1,p_2,\dots,p_N$, a support set $S_{i}$ is computed for each passage $p_i$ using:
\begin{equation}
  S_i \triangleq \{s \in I \cup K : sim(s, p_i) > \varepsilon_i \wedge s \neq p_i\},
\label{eq:kpsupportset}
\end{equation}
for $i=1,\dots, N+M$. Passages are ranked excluding the key phrases $K$ ({\em artificial passages}) according to:
\begin{equation}
\operatorname*{arg\,max}_{s \in (\cup^{n}_{i=1}S_i)-K} \big|\{S_i: s \in S_i\}\big|.
\label{eq:kpmodel:centrality}
\end{equation}

\section{Secure Binary Embeddings}
\label{sec:SBE}
A Secure Binary Embedding (SBE) is a scheme that converts real-valued vectors to bit sequences using band-quan-tized random projections. These bit sequences, which we will refer to as {\em hashes}, possess an interesting property: if the Euclidean distance between two vectors is lower than a threshold, then the Hamming distance between their hashes is proportional to the Euclidean distance between the vectors; if it is higher, then the hashes provide no information about the true distance between the two vectors. This scheme relies on the concept of Universal Quantization \cite{UQ}, which redefines scalar quantization by forcing the quantization function to have non-contiguous quantization regions.

Given an $L$-dimensional vector ${\bf{x}} \in \mathbb{R}^{L}$, the universal quantization process converts it to an $M$-bit binary sequence, where the $m$-th bit is given by
\begin{equation}
q_{m}({\bf x}) = Q \left ( \frac{\left < {\bf x}, {\bf a}_{m} \right > + w_{m}}{\Delta} \right ). \label{eq:scalar_sbe}
\end{equation}
Here $\left<,\right>$ represents a dot product. ${\bf a}_{m} \in \mathbb{R}^{L}$ is a projection vector comprising $L$ i.i.d. samples drawn from $\mathcal{N}(\mu=0,\sigma^2)$, $\Delta$ is a precision parameter, and $w_m$ is a random dither drawn from a uniform distribution over $[0,\Delta]$. $Q(\cdot)$ is a quantization function given by $Q(x) = \lfloor x \text{ mod } 2 \rfloor$. We can represent the complete quantization into $M$ bits compactly in vector form:
\begin{equation}
{\bf q}({\bf x}) = Q \left ( {\bf \Delta}^{-1} ( {\bf Ax}+{\bf w}) \right ), \label{eq:universal_quantizer}
\end{equation}
where ${\bf q}({\bf x})$ is an $M$-bit binary vector, which we will refer to as the {\em hash} of ${\bf x}$. ${\bf A} \in \mathbb{R}^{\it{M}\times \it{L}}$ is a matrix composed of the row vectors ${\bf a}_m$, ${\bf \Delta}$ is a diagonal matrix with entries $\Delta$, and ${\bf w} \in \mathbb{R}^{\it M}$ is a vector composed from the dither values $w_m$.

\begin{figure}[!t]
\centering
\includegraphics[width=\columnwidth]{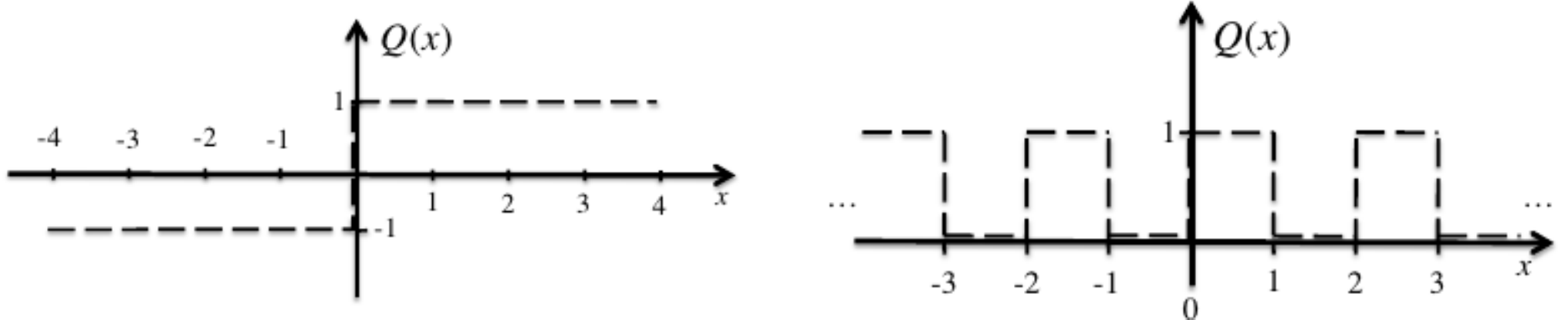}
\caption{1-bit quantization functions.}
\label{figure:1bit_quantization}
\end{figure}

The universal 1-bit quantizer of Equation \ref{eq:scalar_sbe} maps the real line onto $1/0$ in a banded manner, where each band is $\Delta_m$ wide. Figure \ref{figure:1bit_quantization} compares conventional scalar 1-bit quantization (left panel) with the equivalent universal 1-bit quantization (right panel).

The binary hash generated by the Universal Quantizer of Equation \ref{eq:universal_quantizer} has the following properties \cite{SBE}: the probability that the $i^{\rm th}$ bits, $q_i({\bf x})$ and $q_i({\bf x'})$ respectively, of hashes of two vectors ${\bf x}$ and ${\bf x'}$ are identical depends only on the Euclidean distance $d_{E}=\|{\bf x}-{\bf x'}\|$ between the vectors and not on their actual values. As a consequence, the following relationship can be shown \cite{SBE}: given any two vectors ${\bf x}$ and ${\bf x'}$ with a Euclidean distance $d_{E}$, with probability at most $e^{-2t^2M}$ the normalized (per-bit) Hamming distance $d_{H}({\bf q}({\bf x}),{\bf q}({\bf x'}))$ between the hashes of ${\bf x}$ and ${\bf x'}$ is bounded by:

\begin{equation}
\frac{1}{2}-\frac{1}{2}e^{-\left(\frac{\pi \sigma d_{E}}{\sqrt{2}\Delta}\right)^2} - t \leq d_H({\bf q}({\bf x}),{\bf q}({\bf x'})) \leq\frac{1}{2}-\frac{4}{\pi^2}e^{-\left(\frac{\pi \sigma d_{E}}{\sqrt{2}\Delta}\right)^2} + t,
\end{equation}

where $t$ is the control factor. The above bound means that the Hamming distance $d_H({\bf q}({\bf x}),{\bf q}({\bf x'}))$ is correlated to the Euclidean distance $d_{E}$ between the two vectors, if $d_{E}$ is lower than a threshold (which depends on $\Delta$). Specifically, for small $d_{E}$, $E[d_H({\bf q}({\bf x}),{\bf q}({\bf x'}))]$, the expected Hamming distance, can be shown to be bounded by $\sqrt{2\pi^{-1}}\sigma\Delta^{-1} d_{E}$, which is linear in $d_{E}$. However, if the distance between ${\bf x}$ and ${\bf x'}$ is higher than this threshold, then $d_H({\bf q}(\bf x),{\bf q}({\bf x'}))$ is bounded by $0.5-4\pi^{-2}exp\left( -0.5\pi^2\sigma^2 \Delta^{-2} d_{E}^2\right)$, which rapidly converges to $0.5$ and effectively gives us no information whatsoever about the true distance between ${\bf x}$ and ${\bf x'}$.

\begin{figure}[!t]
\centering
\includegraphics[width=0.49\columnwidth]{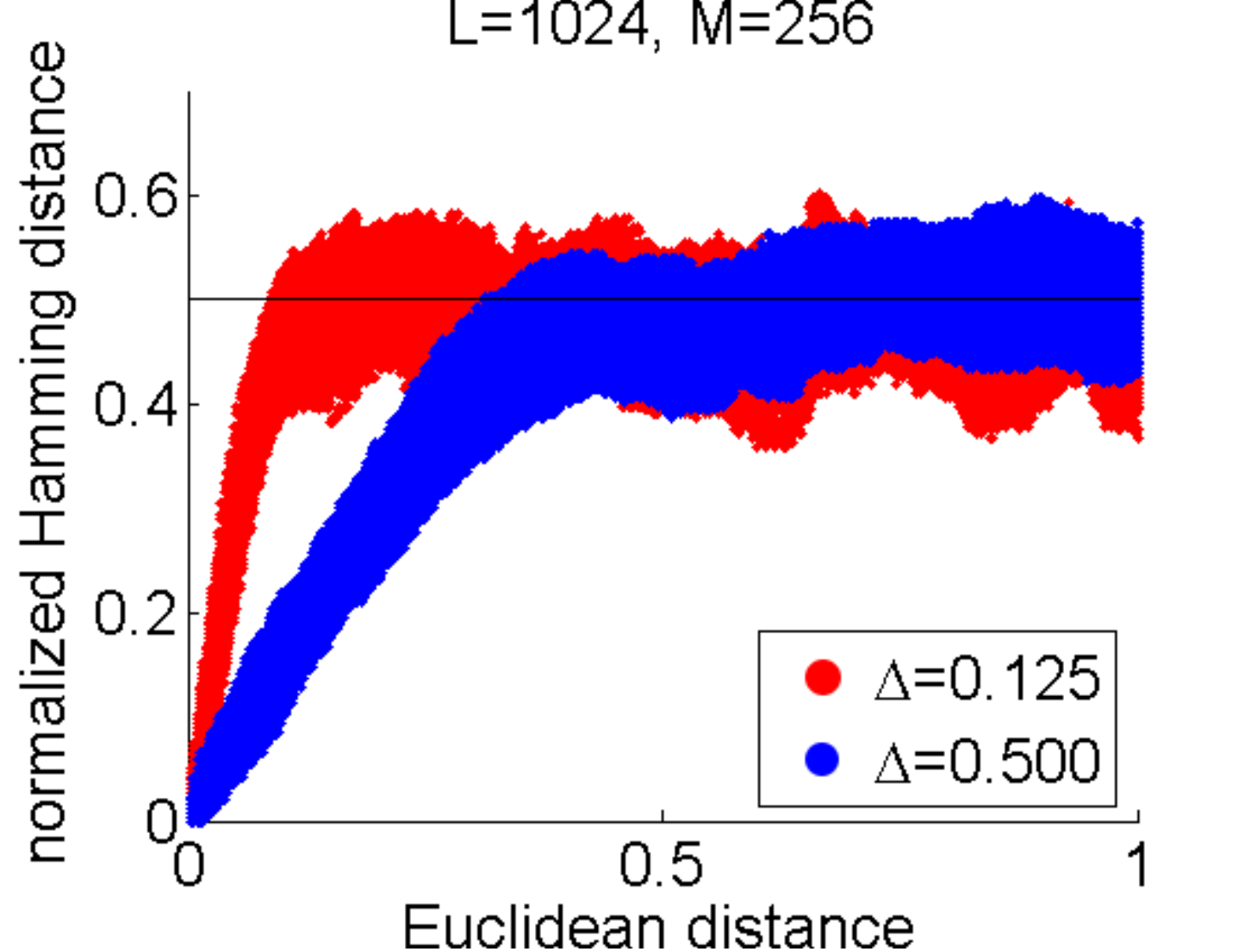}
\includegraphics[width=0.49\columnwidth]{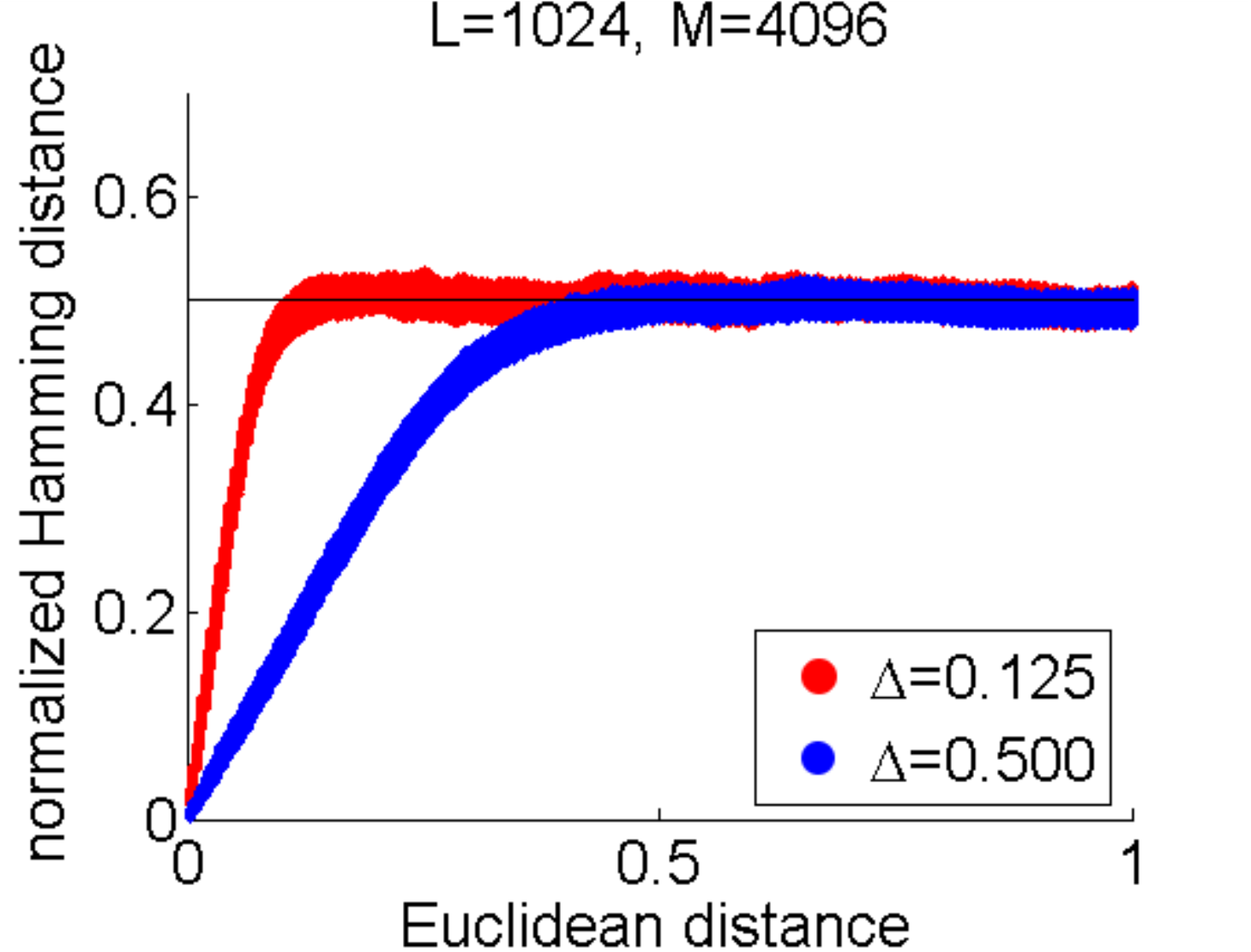}
\caption{SBE behavior as a function of $\Delta$, for two values of $M$.}
\label{figure:sbe}
\end{figure}

In order to illustrate how this scheme works, we randomly generated pairs of vectors in a high-dimensional space ($L=1024$) and plotted the normalized Hamming distance between their hashes against the Euclidean distance between them (Figure \ref{figure:sbe}). The number of bits in the hash is also shown in the figures.

We note that in all cases, once the normalized distance exceeds $\Delta$, the Hamming distance between the hashes of two vectors ceases to provide any information about the true distance between the vectors. We will find this property useful in developing our privacy-preserving system. Changing the value of the precision parameter $\Delta$ allows us to adjust the distance threshold until which the Hamming distance is informative. Increasing the number of bits $M$ leads to a reduction of the variance of the Hamming distance. A converse property of the embeddings is that for all ${\bf x'}$ except those that lie within a small radius of any ${\bf x}$, $d_H({\bf q}({\bf x}),{\bf q}({\bf x'}))$ provides little information about how close ${\bf x'}$ is to ${\bf x}$. It can be shown that the embedding provides information theoretic security beyond this radius, if the embedding parameters ${\bf A}$ and ${\bf w}$ are unknown to the potential eavesdropper. Any algorithm attempting to recover a signal ${\bf x}$ from its embedding ${\bf q}({\bf x})$ or to infer anything about the relationship between two signals sufficiently far apart using only their embeddings will fail to do so. Furthermore, even in the case where ${\bf A}$ and ${\bf w}$ are known, it seems computationally intractable to derive ${\bf x}$ from ${\bf q}({\bf x})$ unless one can guess a starting point very close to ${\bf x}$. In effect, it is infeasible to invert the SBE without strong {\em a priori} assumptions about ${\bf x}$.

\section{Secure Important Passage \\ Retrieval}
\label{sec:secure_passage_retrieval}

\begin{figure*}[!t]
\centering
\includegraphics[width=2.1\columnwidth]{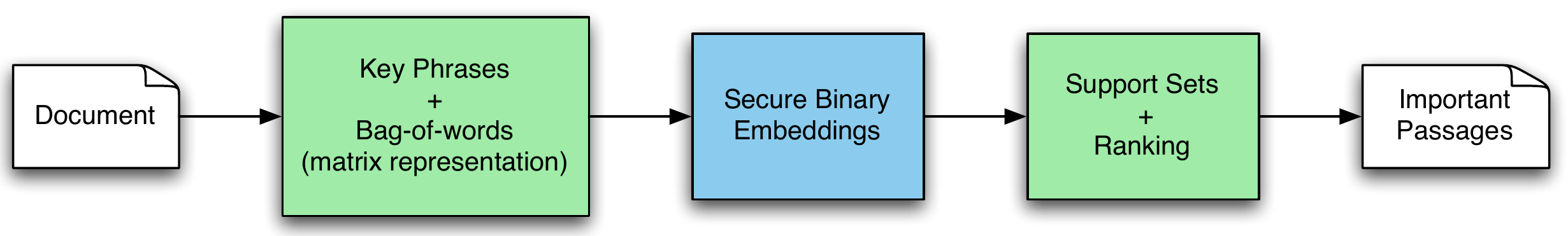}
\caption{Flowchart of the Secure Important Passage Retrieval method.}
\label{fig:KP-Centrality_Flowchart}
\end{figure*}

Our approach for a privacy-preserving important passage retrieval system closely follows the formulation presented in Section \ref{sec:IPR}, and it is illustrated in Figure \ref{fig:KP-Centrality_Flowchart}. 
However, this is a very important difference in terms of who performs each of the steps. Typically there is only one party involved, Alice, who owns the original documents, performs key phrase extraction, combines them with the bag-of-words model in a compact matrix representation, computes the support sets for each documents and finally uses to retrieve the important passages. In our scenario, Alice does not know how to extract the important passages from them or does not possess the computational power to do so. Therefore, she must outsource the retrieval process to a third-party, Bob, who has these capabilities. However, Alice must first obfuscate the information contained in the compact matrix representation. If Bob receives this information as is, he could use the term frequencies to infer on the contents of the original documents and gain access to private or classified information Alice does not wish to disclosure to anyone. Alice computes binary hashes of her compact matrix representation using the method described in Section \ref{sec:SBE}, keeping the randomization parameters ${\bf A}$ and ${\bf w}$ to herself. She sends these hashes to Bob, who computes the support sets and extracts the important passages. Because Bob receives binary hashes instead of the original matrix representation, he must use the normalized Hamming distance instead of the cosine distance in this step, since it is the metric the SBE hashes best relate to. Finally, we returns the hashes corresponding to the important passages to Alice, who then uses them to get the information she desires. 

\section{Experiments}
\label{sec:experiments}
In this section we illustrate the performance of our privacy-preserving approach to Important Passage Retrieval and how it compares to its non-private counterpart. We start by presenting the datasets we used in our experiments, then we describe the experimental setup and finally we present some results.

\subsection{Datasets}
In order to evaluate our approach, we performed experiments on the English version of the Concisus dataset \cite{saggion2012concisus} and the Portuguese Broadcast News (PT BN) dataset \cite{ribeiro:matos:2013}. The Concisus dataset is composed by seventy eight event reports and respective summaries, distributed across three different types of events: aviation accidents, earthquakes, and train accidents. This corpus also contains comparable data in Spanish. However, since our Automatic Key Phrase Extraction (AKE) system uses some language-dependent features, we opted for not using in this part of the dataset in previous work~\cite{Ribeiro&Marujo:2013} nor in this one.

The PT BN dataset consists of automatic transcriptions of eighteen broadcast news stories in European Portuguese, which are part of a news program. News stories cover several generic topics like society, politics and sports, among others. For each news story, there is a human-produced abstract, used as reference.



\subsection{Setup}
We extracted key phrases from both datasets using the MAUI toolkit \cite{Medelyan2010} expanded with shallow semantic features, such as number of named entities, part-of-speech tags and four n-gram domain model probabilities. This expanded feature set leads to improvements regarding the quality of the key phrases. Regarding the Concisus dataset, we extracted yet additional features, such as the detection of rhetorical devices, which further improved the key phrase extraction process \cite{Marujo_LREC_2012}. As for the PT BN dataset, we only used the shallow semantic features as the remaining features were not available \cite{Marujo_Interspeech_2011}.

\subsection{Results}
We present some baseline experiments in order to obtain reference values for our approach. We generated three passages summaries for Concisus Dataset, which are commonly found in online news web sites like Google News. In the experiments using the PT BN dataset, the summary size was determined by the size of the reference human summaries, which consisted in about 10\% of the input news story. For both experiments, we used the Cosine and the Euclidean distance as evaluation metrics, since the first is the usual metric for computing textual similarity, but the second is the one that relates to the Secured Binary Embeddings technique. All results are presented in terms of ROUGE \cite{lin:2004}, in particular ROUGE-1, which is the most widely used evaluation measure for this scenario. The results we obtained for the Concisus and the PT BN datasets are presented in Tables \ref{table:Default_KP_CentralityResults_Concisus} and \ref{table:Default_KP_CentralityResults_PT_BN}, respectively.

\begin{table}[!t]
\center
\begin{tabular}{l|c}
Metric             & ROUGE-1 \\
\hline
Cosine distance    & 0.575   \\
Euclidean distance & 0.507   \\
\end{tabular}
\caption{KP-Centrality results with 40 key phrases using the Concisus dataset.}
\label{table:Default_KP_CentralityResults_Concisus}
\end{table}

\begin{table}[!t]
\center
\begin{tabular}{l|c}
Metric             & ROUGE-1 \\
\hline
Cosine distance    & 0.612   \\
Euclidean distance & 0.590   \\
\end{tabular}
\caption{KP-Centrality results with 40 key phrases using the Portuguese Broadcast News dataset.}
\label{table:Default_KP_CentralityResults_PT_BN}
\end{table}

We considered forty key phrases in our experiments since it is the usual choice when news articles are considered \cite{Marujo_LREC_2012}. As expected, we notice some slight degradation when the Euclidean distance is considered, but we still achieve better results than other state-of-the-art methods such as default centrality \cite{ribeiro:matos:2011} and LexRank \cite{erkan:radev:2004}. Reported results in the literature include $\text{ROUGE-1}=0.443$ and $0.531$ using default Centrality and $\text{ROUGE-1}=0.428$ and $0.471$ using LexRank for the Concisus and PT BN datasets, respectively \cite{Ribeiro&Marujo:2013}. This means that the forced change of metric due to the intrinsic properties of SBE does not affect the validity of our approach in any way.

For our privacy-preserving approach we performed experiments using different values for the SBE parameters. The results we obtained in terms of ROUGE for the Concisus and the PT BN datasets are presented in Tables \ref{table:KP-CentralityResults_with_privacy_Concisus} and \ref{table:KP-CentralityResults_with privacy_PT_BN}, respectively.
\begin{table}[!t]
\center
\begin{tabular}{l|ccccc}
leakage      & $\sim5\%$ & $\sim25\%$ & $\sim50\%$ & $\sim75\%$ & $\sim95\%$ \\
\hline
{\em bpc}=4  & 0.365     & 0.437      & 0.465      & 0.486      & 0.495      \\
{\em bpc}=8  & 0.424     & 0.384      & 0.436      & 0.452      & 0.500      \\
{\em bpc}=16 & 0.384     & 0.416      & 0.450      & 0.463      & 0.517      \\
\end{tabular}
\caption{KP-Centrality using SBE and the Concisus dataset, in terms of ROUGE-1.}
\label{table:KP-CentralityResults_with_privacy_Concisus}
\end{table}
\begin{table}[!t]
\center
\begin{tabular}{l|ccccc}
leakage      & $\sim5\%$ & $\sim25\%$ & $\sim50\%$ & $\sim75\%$ & $\sim95\%$ \\
\hline
{\em bpc}=4  & 0.314     & 0.340      & 0.470      & 0.478      & 0.562      \\
{\em bpc}=8  & 0.327     & 0.324      & 0.486      & 0.507      & 0.527      \\
{\em bpc}=16 & 0.338     & 0.336      & 0.520      & 0.473      & 0.550      \\
\end{tabular}
\caption{KP-Centrality using SBE and the Portuguese Broadcast News dataset, in terms of ROUGE-1.}
\label{table:KP-CentralityResults_with privacy_PT_BN}
\end{table}
Leakage refers to the fraction of SBE hashes for which the normalized Hamming distance $d_{H}$ is proportional to the Euclidean distance $d_{E}$ between the original data vectors. The amount of leakage is exclusively controlled by $\Delta$. Bits per coefficient ($bpc$) is the ratio between the number of measurements $M$ and the dimensionality of the original data vectors $L$, i.e., $bpc=M/L$. As expected, increasing the amount of leakage (i.e. increasing $\Delta$) leads to improving the retrieval results. Surprisingly, changing the values of $bpc$ does not lead to improved performance. The reason for this results might be due to the KP-Centrality method using support sets that consider multiple partial representations of the documents. Nevertheless, the most significant results is that for $95\%$ leakage there is an almost negligible loss of performance. This scenario, however, does not violate our privacy requisites in any way, since although most of the distances between hashes are known, there is no way to use this information to learn anything about the original underlying information.

%
%
%
\section{Conclusions and Future Work}
\label{sec:conclusions}
In this work, we introduced a privacy-preserving technique for performing Important Passage Retrieval that performs similarly to their non-private counterpart. Our Secure Binary Embeddings based approach provides secure document representations that allows for sensitive information to be processed by third parties without any risk of sensitive information disclosure. Although there was some slight degradation of results regarding the baseline, our approach still outperforms other state-of-the-art methods like default Centrality and LexRank, but with important advantage that no private or classified information is disclosed to third parties.

For future work we intend to use the secure representation based on Secure Binary Embeddings in multi-document important passage retrieval. Another additional research line that we would like to purse is to apply this privacy preserving technique in other Information Retrieval tasks, such as classified military and medical records retrieval.

\section{Acknowledgments}
We would like to thank FCT for supporting this research through PPEst-OE/EEI/LA0021/2013, the Carnegie Mellon Portugal Program, PTDC/EIA-CCO/122542/2010, and grants SFRH/BD/33769/2009 and SFRH/BD/71349/2010. We would like to thank NSF for supporting this research through grant 1017256.

%
\bibliographystyle{abbrv}
\bibliography{sigirrsp098}  
%
%
%
\end{document}